\documentclass{mem}
\usepackage{natbib}\usepackage{txfonts}\usepackage{balance}
\usepackage{graphicx}
\usepackage[a4paper]{hyperref}
\idline{75}{282}
\begin{document}
\def\teff{$T\rm_{eff }$}
\def\kms{$\mathrm {km s}^{-1}$}

\title{
From MAGIC to CTA: the INAF participation to Cherenkov 
Telescopes experiments for Very High Energy Astrophysics }

   \subtitle{}

\author{
L. Angelo \,Antonelli\inst{1} \\
on behalf of the INAF MAGIC collaboration
          }

  \offprints{antonelli@oa-roma.inaf.it}

\institute{
Istituto Nazionale di Astrofisica --
Osservatorio Astronomico di Roma, Via Frascati, 33,
I-00040 Monte Porzio Catone (RM), Italy;
\email{a.antonelli@oa-roma.inaf.it}
}

\authorrunning{L.A. Antonelli}

\titlerunning{The MAGIC and CTA experiments}

\abstract{
The next decade can be considered the "golden age" of the Gamma Ray Astronomy with the two satellites for Gamma Ray Astronomy (AGILE and GLAST) in orbit. Therefore, thanks to many other X-ray experiments already in orbit (e.g. Swift, Chandra, NewtonXMM, etc.) it will be possible to image the Universe for the first time all over the electromagnetic spectrum almost contemporarily. The new generations of ground-based very high gamma-ray instruments are ready to extend the observed band also to the very high frequencies. Scientists from the Italian National Institute for Astrophysics (INAF) are involved in many, both space- and ground- based gamma ray experiments, and recently such an involvement has been largely improved in the field of the Imaging Atmospheric Cherenkov Telescopes (IACT).  INAF is now member of the MAGIC collaboration and is participating to the realization of the second MAGIC telescope. MAGIC, as well other IACT experiments, is not operated as an observatory so a proper guest observer program does not exist. A consortium of European scientists (including INAF scientists) is thus now thinking to the design of  a new research infrastructure: the Cherenkov Telescope Array (CTA). CTA is conceived to provide 10 times the sensitivity of current instruments, combined with increased flexibility and increased coverage from some 10 GeV to some 100 TeV. CTA will be operated as an observatory to serve a wider community of astronomer and astroparticle physicists.  
}
\maketitle{}

\section{Introduction}
The results of the latest generation of ground-based gamma-ray instruments such as H.E.S.S., MAGIC, CANGAROO or VERITAS, have shown that the very high energy gamma-ray astronomy has grown to a genuine branch of astronomy. Cherenkov Telescopes are now allowing imaging, photometry and spectroscopy of sources of high-energy radiation with good sensitivity and good angular resolution. The number of known sources of very high energy gamma rays is continuously growing (now approaching 100), and source types include a lot of different classes of known objects as well as unidentified sources without obvious counterpart. The major scientific objective of $\gamma$-ray astronomy is the understanding of the production, acceleration, transport and reaction mechanisms of VHE particles in astronomical objects. This is tightly linked to the search for sources of the cosmic rays connecting astrophysics with particle physics, so the physics program of the (e.g.) MAGIC telescope includes topics, 
both of fundamental physics and astrophysics. In the next 10 years, GLAST and (for less time) AGILE satellites will observe the Universe in the MeV-GeV band providing a unique opportunity for Cherenkov Telescopes to observe the same sources and to cross-calibrate instruments in the GeV band.  In such a scenario the development of a next generation of Cherenkov Telescopes is mandatory in order to achieve 10 times the sensitivity of current instruments, an increased flexibility and an increased coverage from some 10 GeV to some 100 TeV. \\
Since 2003 the Italian National Institute for Astrophysics (INAF) is collecting all the italian research institutions operating in the field of Astronomy and Astrophysics (formerly Astronomical Observatories and CNR institutes) and has inherited the long lasting experience in the field of the high energy astrophysics of the former institutes. Recently INAF has decided to join the MAGIC project in the field of VHE gamma astrophysics providing to the project both technological and scientific contribution. 
\begin{figure*}[t!]
\resizebox{\hsize}{!}{\includegraphics[clip=true]{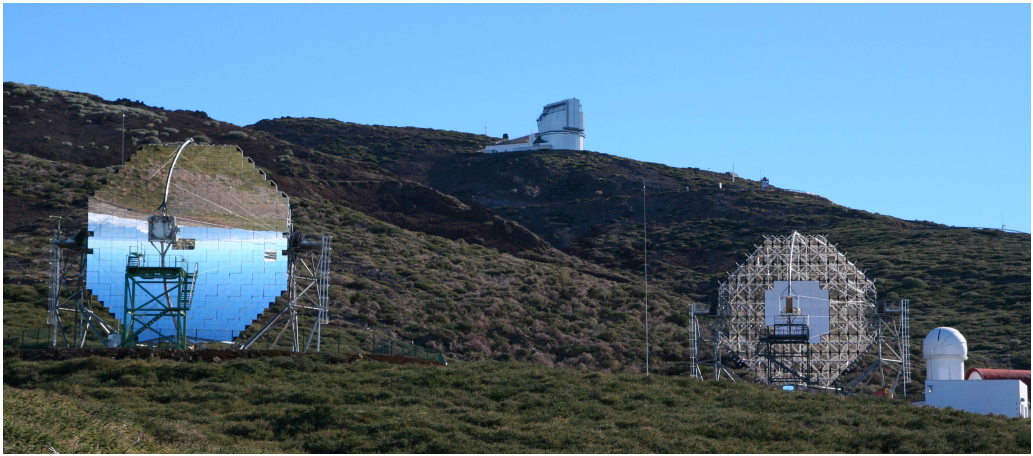}}
\caption{\footnotesize MAGIC and MAGIC 2 (under construction) \label{magic} in La Palma on Dec 2007. On the background the Italian {\it Telescopio Nazionale Galileo}, a 3.5 m optical telescope operated by INAF.
}
\label{eta}
\end{figure*}

\section{The MAGIC Experiment }

MAGIC\footnote{http://wwwmagic.mppmu.mpg.de/} \citep{CortinaICRC} is currently the largest single-dish Imaging Air 
Cherenkov Telescope (IACT) in operation. 
Located on the Canary Island La Palma ($28.8^\circ$N, $17.8^\circ$
W, 2200~m a.s.l.), it has a 17-m diameter tessellated parabolic mirror, supported 
by a light-weight carbon fiber frame. It is equipped with a high-quantum-efficiency 
576-pixel $3.5^\circ$ field-of-view photomultiplier tube (PMT) camera. 


The accessible energy range spans from 40-50 GeV (trigger threshold at small 
zenith angles) up to tens of TeV; the present analysis threshold at zenith is 60 GeV. 
The $5\sigma$ sensitivity of MAGIC is $\sim$~1.5\% of the Crab Nebula flux in 50
hours of observations. The relative energy resolution is about 25\% above 100 GeV 
and about 20\% above 200 GeV.  The $\gamma$ point spread function (PSF) is slightly 
less than $0.1$~degrees \citep{crab}. 
The sensitivity of MAGIC as calculated from the Monte Carlo simulation (MC)  is shown 
together with the expected sensitivity of other gamma-ray detectors in the GeV and 
TeV range in Fig.~2.

\subsection{Science with MAGIC} 

The observation program of MAGIC includes several galactic and extragalactic
types of sources such as Supernova Remnants (SNR), pulsars, microquasars and Active
Galactic Nuclei (AGNs). The low energy threshold allows MAGIC to extend the
observation of extragalactic sources up to z$\sim$1 and beyond. The high sensitivity and 
the low energy threshold of the MAGIC telescope allowed detailed studies of the spectral 
features of these sources, as well as the observation of flux variability on short timescales.
Another unique feature of MAGIC is the fast repositioning time of the telescope that allows 
to observe gamma ray bursts within 20-40~s after the alert by satellite detectors. 
Besides, MAGIC has a huge potential for studies related to fundamental physics e.g. search 
for dark matter,  quantum gravity, etc. 
Since its first cycle of data taking (February 2005), the MAGIC telescope has observed nine 
galactic and twelve extra-galactic sources of VHE $\gamma$-radiation. Ten of these objects 
had never been detected before in VHE $\gamma$-rays. Very important results In the field 
of galactic objects are the discovery of the VHE emission from the binary 
system LS I +61 303~\citep{lsi}, possibly ($4.1 \sigma$ significance after trial correction) 
from Cyg X-1 \citep{MAGIC_CygX1} and recently the detection of the Crab pulsar in very 
high energy gamma rays \citep{pulse}. MAGIC published up to now the detection of 12 
extragalactic sources and all of them are well-established Active Galactic Nuclei (AGN).
Among these there are: the very important observation of 3C279 $(z=0.536)$ \citep{mexico}
putting strong constraints on the Extragalactic Background Light (EBL) models; the observation
of BL Lac the first member of  the ``Low-frequency peaked BL Lacs'' (LBL) ever detected to emit 
in the VHE region\citep{bllac}; the observation of a big flare from Mkn 501\citep{Mkn501} during 
which an unprecedent short doubling time of 2-4 minutes was detected and the rapid increase 
in the flux level was accompanied by a hardening of the differential spectrum. 
 MAGIC has been built specifically also for GRB observations and it is able to automatically 
 repoint and start the observation within a maximum time (depending on the position in the sky)
of 40 seconds. Up to now, many GRBs have been targeted with the MAGIC telescope and upper
limits for the flux were derived for all events \citep{grb}.
\begin{figure}
\begin{center}
\includegraphics[width=\columnwidth]{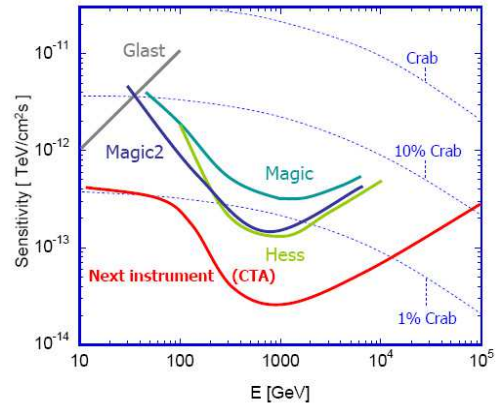}
\end{center}
  \caption{\footnotesize Sensitivities\label{sensi} for some operating and proposed gamma detectors. The GLAST result sums 1 year of data taking, the IACTs sum 50 hours.}
\end{figure}

\subsection{The second MAGIC telescope} 

The construction of a second telescope, MAGIC2, close (about 80~m) to the original one, has started.  
It incorporates some minor modifications suggested by the experience of running MAGIC for the last three years, as well as some important changes. In particular larger (1 m$^{2}$ surface) and lighter mirrors have been developed under the responsibility of the INAF using a new technique based on the so called ÒCold slumpingÓ. 
Each mirror is realized by a glass sheet maintained in shape by a master mould and fixed onto a honeycomb structure to produce a panel. Then on the concave side one deposits a reflecting coating (Aluminum) and a thin protective coating (Quartz) obtaining a very high reflectivity and, at the same time, a large durability. This approach is very attractive since massive production with short manufacturing times is possible (e.g. 5 mirrors per day if 5 masters are available). A major improvement of the Data Acquisition System is also under construction to reduce the Night Sky Background contribution to the signal. 

MAGIC2 is expected to be ready by September 2008;  stereo observations will then be operational, allowing an increase in the sensitivity by at least a factor of 3, and other improvements in the energy and direction reconstruction. With the advent of MAGIC2, we will reach a level of 1\% Crab Unit in 50 hours of data taking.  Meanwhile, the AGILE results should come and GLAST should become fully operational, closing 
the current observational gap between $\sim1$ and $60\:\mathrm{GeV}$ and extending observations of the electromagnetic radiation, without breaks, up to almost $100\:\mathrm{TeV}$.
The inauguration of MAGIC2 is scheduled for September 18th, 2008.

\section{The Cherenkov Telescope Array}

The Cherenkov Telescope Array (CTA)\footnote{http://www.mpi-hd.mpg.de/hfm/CTA/} is a European initiative to build up the next generation 
ground-based gamma-ray instrument as an open observatory to serve a wider astrophysics 
community.  The aims of the CTA observatory are to increase sensitivity in the core energy 
range from about 100 GeV to about 10 TeV by roughly one order of magnitude, and to expand 
the energy range for very high energy gamma astronomy towards both lower and higher 
energies, effectively increasing the usable energy coverage by a factor of 10.  The observatory 
should consist of two arrays: a southern hemisphere array, which covers the full energy range from 
some 10 GeV to about 100 TeV to allow for a deep investigation of galactic sources, and of the 
central part of our Galaxy, but also for the observation of extragalactic objects. A northern hemisphere array, consisting of the low energy instrumentation (from some 10 GeV to ~1 TeV) complements the observatory and is dedicated mainly to northern extragalactic objects. The all sky observatory with its two sites will be operated by one single consortium. A significant fraction of the observation time will be open to the general astrophysical community and facilities for user support will be provided. Implementation of first prototype telescope(s) of the system could start in 2010 after a period of a detailed design study and optimization, site evaluation and production of industrial prototypes of components.
CTA is being considered as ÒEmerging ProposalÓ in the 2006 roadmap report of the European Strategy Forum on Research Infrastructures (ESFRI). The construction of CTA as a next-generation facility for ground-based very-high-energy gamma-ray astronomy is Òvery strongly recommendedÓ in the current ASPERA roadmap.
INAF is participating to the CTA project since the beginning and both scientists and engineers from INAF are involved in the different Work Packages of the project. In particular INAF is directly involved in the study and development of mirrors,  detectors and electronics,  data analysis, site selection and observatory operations as well in the CTA science.   

\section{Conclusions}
Thanks to GLAST, AGILE, MAGIC 2, H.E.S.S. 2 and other experiments, the next decade can be considered the "golden age" of the Gamma Ray Astronomy allowing for the first time a deep view of  the gamma-ray Universe. CTA will be the 
new generation of ground-based very high gamma-ray instruments and it will increase the number of the detected sources in the 100 GeV 10 TeV band by a factor of 10. Scientists from INAF are involved in many gamma ray experiments and recently such an involvement has been largely improved in the field of the IACTs with the 
participation to both the MAGIC collaboration and the CTA project.

\begin{acknowledgements}
We thank the INAF Projects Dept. for the support to the 
MAGIC activities. CTA activities are partly supported by 
the PRIN-INAF 06.
\end{acknowledgements}

\bibliographystyle{aa}

\end{document}